\documentclass[
 reprint,
superscriptaddress,
 amsmath,amssymb,
 aps,
]{revtex4-2}

\usepackage{graphicx}
\usepackage{dcolumn}
\usepackage{bm}
\usepackage{lineno}
\usepackage{amsfonts}
\usepackage{makeidx}
\usepackage{colortbl}
\usepackage{multirow}
\usepackage{lmodern}
\usepackage{afterpage}
\usepackage{array}
\usepackage{physics}
\usepackage{upgreek}
\usepackage{xcolor}
\usepackage{xr} 

\usepackage{caption}
\usepackage{cuted}

\newcommand{\ii}{\operatorname{i}}
\newcommand{\ee}{\operatorname{e}}

\usepackage{xcolor}

 \usepackage[english]{babel}

\begin{document}

\preprint{APS/123-QED}

\title{Cross-correlation scheme for quantum optical coherence tomography based on Michelson interferometer}

\author{Anna Romanova}%
    \email{romanova.phys@gmail.com}
    \affiliation{%
Quantum Research Center, Technology Innovation Institute, Masdar City, Abu Dhabi, UAE}
    \affiliation{Institut für Physik, Humboldt-Universität zu Berlin, 12489 Berlin, Germany}%
\author{Vadim Rodimin}%
    \affiliation{%
Quantum Research Center, Technology Innovation Institute, Masdar City, Abu Dhabi, UAE}
\author{Konstantin Katamadze}%
    \affiliation{%
Quantum Research Center, Technology Innovation Institute, Masdar City, Abu Dhabi, UAE}

\date{\today}

\begin{abstract}
Quantum optical coherence tomography (QOCT) offers a simple way to cancel dispersion broadening in a sample while also providing twice the resolution compared to classical OCT. 
However, to achieve these advantages, a bright and broadband source of entangled photon pairs is required.
A simple implementation uses collinear spontaneous parametric down-conversion in a Michelson interferometer (MI), yet this autocorrelation scheme suffers from parasitic terms and sensitivity to phase noise.
Here, we introduce a cross-correlation MI-based QOCT that overcomes these drawbacks, significantly advancing QOCT toward practical applications.

\end{abstract}

\maketitle


\section{Introduction}\label{sec:Introduction}
\begin{figure}[tbh]
    \centering
    \includegraphics[width= \columnwidth]{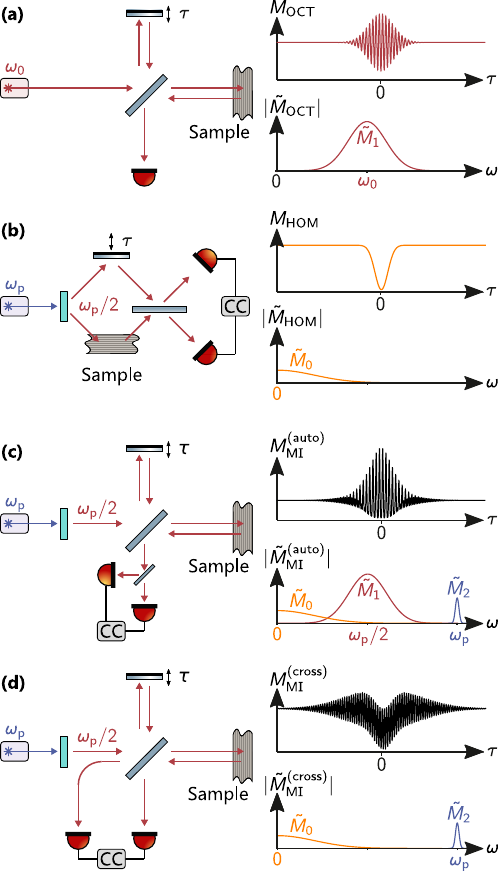}
      \caption{OCT and QOCT techniques: (a) time-domain OCT, (b) traditional HOM-based QOCT, (c) MI-based QOCT with autocorrelation measurements, and (d) the studied scheme of MI-based QOCT with cross-correlation measurements. The left column presents sketches of the experimental schemes, while the right column shows the corresponding PSF interferograms  $M(\tau)$ (from a single reflective layer) and their Fourier transforms $|\tilde{M}(\omega)|$.}
    \label{fig:Schemes}
\end{figure}

The resolution of far-field imaging systems is limited by the optical radiation wavelength and the numerical aperture (NA). 
While the transverse resolution is inversely proportional to NA, the axial resolution is inversely proportional to the square of NA~\cite{pawley_handbook_2006}. 
This limitation is particularly critical in applications where NA is significantly limited, such as retinal imaging through an eye pupil~\cite{drexler_optical_2015}.

The technique that addresses this limitation is optical coherence tomography (OCT), introduced in Ref.~\cite{huang_optical_1991} and widely described in Ref.~\cite{drexler_optical_2015}.
In the time-domain OCT scheme, presented in Fig.~\ref{fig:Schemes}a, the sample is placed in one arm of a Michelson interferometer (MI), while the other arm's delay, $\tau$, is varied. 
Each reflective layer of the sample generates an interferogram $M_\text{OCT}(\tau)$, allowing one to reconstruct the structure of the sample -- coordinates and reflectivities of the layers.  
The interferogram obtained from a single layer can be considered as a point-spread function (PSF) of the imaging system. 
Its width determines the axial resolution, independent of NA but limited by the coherence length of the optical source. 
Thus, a low-coherence broadband source is required for fine axial resolution. 
However, any dispersion in the sample or setup components broadens the interferogram, severely limiting the scanning depth of OCT unless specific sample dispersion properties are known and compensated.

Several dispersion‐cancellation strategies for OCT have been demonstrated.  Some -- like the hardware phase‐compensation method~\cite{hitzenberger_dispersion_1999} -- require precise a~priori knowledge of the sample’s dispersion and only cancel dispersion at a single imaging depth.  Others rely on elaborate nonlinear‐optics implementations, for example phase‐conjugation or chirped‐pulse schemes~\cite{erkmen_phaseconjugate_2006, kaltenbaek_quantuminspired_2008, legouet_experimental_2010, lavoie_quantumoptical_2009}, which demand high‐power lasers and complex setups. 
Other dispersion-cancellation approaches rely on intensive numerical spectral post-processing \cite{banaszek_blind_2007, resch_classical_2007, ryczkowski_experimental_2016, shirai_intensityinterferometric_2014, jensen_alldepth_2018, liu_dispersion_2020}, which makes them intrinsically sensitive to interferometric phase and timing instabilities.

Quantum optical coherence tomography (QOCT)~\cite{abouraddy_quantum-optical_2002, Nasr2003}, based on Hong-Ou-Mandel (HOM) interference~\cite{Hong1987b}, presented in Fig.~\ref{fig:Schemes}b, overcomes the dispersion broadening issue and also provides double resolution.
In QOCT, each reflective layer produces a dip $M_\text{HOM}(\tau)$, where the width, inversely proportional to the bandwidth of the spontaneous parametric down-conversion (SPDC)~\cite{Klyshko1988a} field, defines the system's axial resolution. Notably, QOCT is insensitive to group-velocity dispersion if the SPDC pump field is sufficiently narrowband~\cite{okanoDispersionCancellationHighresolution2013}, and it provides twice the axial resolution of OCT with the same source spectrum. 
Additionally, as mentioned in Ref.~\cite{Hong1987b}, HOM interference does not require any phase stabilization or sub-wavelength scanning steps, and the interferogram pattern $M_\text{HOM}(\tau)$ is much clearer and easier to process compared to $M_\text{OCT}(\tau)$.

Despite its advantages, the practical application of QOCT remains limited due to the long measurement times required by the low brightness of broadband SPDC sources. Bright sources are more effortlessly produced via collinear phase matching, which allows for effective use of longer crystal lengths. As mentioned in Ref.~\cite{Katamadze2022}, for broadband sources, degenerate phase matching types-I or 0 are preferable, though this leads to both photons being in the same optical mode, preventing their deterministic separation for HOM interference.

Recent studies suggest that QOCT can also be implemented using a brighter, broader~\cite{baek_SpectralProperties2008}, and more adjustable collinear type-0/type-I SPDC source coupled to an MI with an autocorrelation measurement scheme~\cite{odateTwophotonQuantumInterference2005, lopez-mago_coherence_2012, lopez-mago_quantum-optical_2012, yoshizawa_telecom-band_2014, katamadze_broadband_2025}. 
In this configuration (Fig.~\ref{fig:Schemes}c), the interferometer's output signal is split, and the coincidence count rate interferogram $M_\text{MI}^\text{(auto)}(\tau)$, is measured.

The coincidence interferogram $M_\text{MI}^\text{(auto)}(\tau)$ has a complex structure.
In addition to the desired HOM term \(M_0\), it contains several parasitic contributions.  
Its Fourier transform \(\tilde{M}_\mathrm{MI}^\mathrm{(auto)}(\omega)\) reveals that these terms occupy distinct central frequencies, allowing \(M_0\) to be extracted by post-processing via Fourier filtering.

However, achieving sub-micrometer axial resolution~\cite{katamadze_broadband_2025} requires an ultra-broadband SPDC source, which causes overlap between the target and parasitic components in the Fourier domain.  
Phase instabilities further broaden these features, making post-processing more challenging, reducing the usable spectral bandwidth, and ultimately limiting the axial resolution.

To overcome these limitations, we explore and experimentally demonstrate a \emph{cross-correlation} MI-based QOCT scheme (Fig.~\ref{fig:Schemes}d).  
This scheme combines dispersion cancellation, doubled resolution, and robustness to phase instability of the standard HOM approach with the high brightness of collinear SPDC, while eliminating overlap between parasitic Fourier components of the interferogram.  
By monitoring coincidences between the MI’s two output ports (one of which is recycled to the input via a circulator), we obtain the PSF interferogram \(M_\mathrm{MI}^\mathrm{(cross)}(\tau)\).  
Its Fourier transform \(\tilde{M}_\mathrm{MI}^\mathrm{(cross)}(\omega)\) contains only a narrow pump-frequency peak $\tilde M_2$ and the target HOM term $\tilde M_0$.  
Because the parasitic peak is well separated from the HOM dip, it can be removed by simple Fourier filtering or by averaging over high-frequency oscillations during a coarse mechanical scan -- without limiting SPDC bandwidth or requiring sub-wavelength path control.

The cross-correlation approach was recently demonstrated for high-precision refractive-index measurements in optical fibers \cite{reisner_quantumlimited_2022}, however, in that application the PSF width is not critical because each scan measures a single path length and the peak can be located with sub-PSF precision.  By contrast, here we provide a comprehensive study of a Michelson-based cross-correlation QOCT: we compare its resolution and signal-to-noise ratio (SNR) with both autocorrelation QOCT and classical OCT, analyze how the axial resolution depends on the source spectrum, demonstrate enhanced robustness to phase noise and optical dispersion, and finally apply the method to defectoscopy -- a task where PSF narrowing and dispersion cancellation are essential.

The paper is organized as follows. Section~\ref{sec:Theory} presents the theoretical description of classical OCT and of both autocorrelation and cross-correlation MI-based QOCT schemes and highlights the advantages of the latter. Section~\ref{sec:Experiment} describes the experimental setup that implements both QOCT schemes alongside a classical OCT reference. Section~\ref{sec:results} compares resolution and SNR across the three techniques and includes a defectoscopy demonstration.

\section{Theory}\label{sec:Theory}

\begin{figure}
    \centering
    \includegraphics[width=\columnwidth]{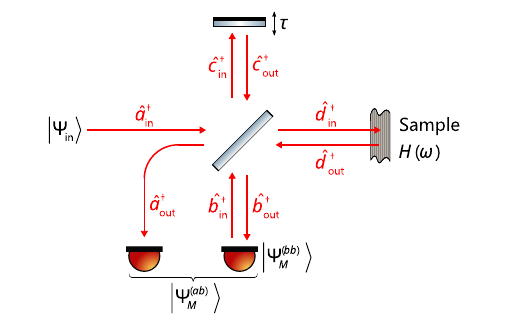}
    \caption{Michelson interferometer input and output modes, input state \(\ket{\Psi_\text{in}}\) and measurement projections for autocorrelation \(\ket{\Psi_M^{(bb)}}\) and cross-correlation \(\ket{\Psi_M^{(ab)}}\).}
    \label{fig:Michelson}
\end{figure}

In this section we follow the approach of Ref.~\cite{katamadze_broadband_2025} to describe quantum interference in the MI-based QOCT schemes of Fig.~\ref{fig:Schemes}a, c, and d and to compare them theoretically.

Consider interference of a photon pair in MI. 
The interferometer scheme is illustrated in Fig.~\ref{fig:Michelson}.
It consists of a 50/50 beam splitter, a reflective sample, and a translated mirror in a reference channel. 
It connects four input and four output optical modes. 
Photon creation operators in these modes are denoted as $\hat{a}^\dagger_{\text{in}}$, $\hat{a}^\dagger_{\text{out}}$, $\hat{b}^\dagger_{\text{in}}$, $\hat{b}^\dagger_{\text{out}}$, $\hat{c}^\dagger_{\text{in}}$, $\hat{c}^\dagger_{\text{out}}$, $\hat{d}^\dagger_{\text{in}}$, $\hat{d}^\dagger_{\text{out}}$. The beam splitter connects these operators as follows:

\begin{equation}
\begin{aligned}\label{eq:Michelson1.matrices1}
   \left( \begin{matrix}
   \hat{a}_{\text{in}}^{\dagger }  \\
   \hat{b}_{\text{in}}^{\dagger }  \\
\end{matrix} \right)&=\frac{1}{\sqrt{2}}\left( \begin{matrix}
   1 & \text{i}  \\
   \text{i} & 1  \\
\end{matrix} \right)\left( \begin{matrix}
   \hat{d}_{\text{in}}^{\dagger }  \\
   \hat{c}_{\text{in}}^{\dagger }  \\
\end{matrix} \right)
,\\
\left( \begin{matrix}
   \hat{c}_{\text{out}}^{\dagger }  \\
   \hat{d}_{\text{out}}^{\dagger }  \\
\end{matrix} \right)&=\frac{1}{\sqrt{2}}\left( \begin{matrix}
   1 & -\text{i}  \\
   -\text{i} & 1  \\
\end{matrix} \right)\left( \begin{matrix}
   \hat{a}_{\text{out}}^{\dagger }  \\
   \hat{b}_{\text{out}}^{\dagger }  \\
\end{matrix} \right).
\end{aligned}
\end{equation}
Let $H(\omega)$ denote the sample response function and $\tau$ the delay in the reference channel.
Then input and output modes are connected as follows:
\begin{equation}\label{eq:Michelson1.matrices2}
    \left( \begin{matrix}
   \hat{c}_{\text{in}}^{\dagger }  \\
   \hat{d}_{\text{in}}^{\dagger }  \\
\end{matrix} \right)=\left( \begin{matrix}
   {{\operatorname{e}}^{\operatorname{i}\omega \tau }} & \text{0}  \\
   \text{0} & H\left( \omega  \right)  \\
\end{matrix} \right)\left( \begin{matrix}
   \hat{c}_{\text{out}}^{\dagger }  \\
   \hat{d}_{\text{out}}^{\dagger }  \\
\end{matrix} \right).
\end{equation}

The general description of the single-photon and photon pair MI-interference is presented in {Appendix}~\ref{appendix:MIgeneral}. 
Here we make two assumptions.
\begin{itemize}
    \item The sample has a single layer without dispersion, with the response function
\begin{equation}\label{eq:single}
    H(\omega) = r\,\ee^{\ii \omega T}, \quad r^{2} \equiv R,
\end{equation}
where $R$ is the layer reflectivity.
Here, $\omega T = k d = \tfrac{n \omega d}{c}$, with $d$ the layer thickness, $n$ the refractive index, and $c$ the speed of light.  
Thus, $T = \tfrac{nd}{c}$.

\item The input state consists of photons with the same polarization, propagating in input mode $a$ and having Gaussian spectral distribution.
\end{itemize}

First, we consider a classical or single-photon state with a Gaussian spectral distribution
\begin{equation}
\label{eq:gaussian}
   G(\omega|\omega_0, \sigma)\equiv \frac{1}{\sqrt{2 \pi}\sigma}\exp\left[-\frac{(\omega-\omega_0)^2}{2\sigma^2}\right]
\end{equation}
with the  central frequency $\omega_0$ and the bandwidth $\sigma$ (in this section we define the bandwidth as a standard deviation of the Gaussian distribution). Measuring the output intensity (count rate) in the mode $b$ as a function of $\tau$ yields the following interferogram:

\begin{equation}
\label{eq:classical}
   M^{(b)}(\tau)= \frac{1+R}{4}-\frac{r}{2}\exp \left[ -\frac{{{\left( T-\tau  \right)}^{2}}}{2 \sigma^{-2}} \right]\cos \left[ {{{\omega }_{0}}\left( T-\tau  \right)} \right],
\end{equation}
presented in Fig.~\ref{fig:Schemes}a (see Appendix~\ref{appendix:MIgeneral} for details). This interferogram represents the PSF of classical OCT.

\begin{figure}
    \centering
    \includegraphics[width= \columnwidth]{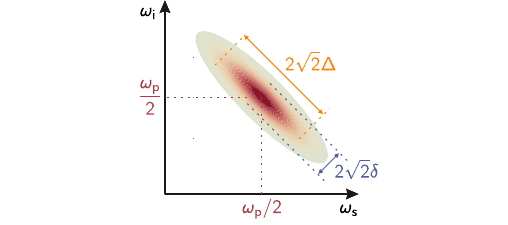}
    \caption{Two-photon spectral distribution $\abs{f(\omega_{\text{s}},\omega_{\text{i}})}$ for degenerate case.}
    \label{fig:TwoOmega}
\end{figure}

For QOCT we consider a collinear degenerate SPDC state with type-0/type-I phase matching:
\begin{equation}\label{eq:Michelson1.Psiin}
    \left| {{\Psi }_{\text{in}}} \right\rangle =
    \frac{1}{\sqrt{2}}
    \iint{\text{d}{{\omega }_\text{s}}\text{d}{{\omega }_\text{i}}}F\left( {{\omega }_\text{s}},{{\omega }_\text{i}} \right)\hat{a}_{\text{in}}^{\dagger }\left( {{\omega }_\text{s}} \right)\hat{a}_{\text{in}}^{\dagger }\left( {{\omega }_\text{i}} \right)\left| \text{vac} \right\rangle, 
\end{equation}
where ${F}({{\omega }_\text{s}},{{\omega }_\text{i}})$ is a spectral amplitude and
\begin{equation}
\label{eq:degenerate}
   {\left| F({{\omega }_\text{s}},{{\omega }_\text{i}}) \right|}^{2}= 2
   \underbrace{G(\omega_\text{s}-\omega_\text{i}|0, \Delta)}_\text{phase matching}\times \underbrace{G(\omega_\text{s}+\omega_\text{i}|\omega_\text{p}, \delta)}_\text{pump}
\end{equation}
is a Gaussian approximation of the photon-pair spectral distribution~\cite{Mikhailova2008,Fedorov2009}, visualized in Fig.~\ref{fig:TwoOmega}.
Here $\Delta$ represents the phase matching bandwidth and $\delta \ll \Delta$ is the  pump bandwidth.

The corresponding marginal single-photon spectral distribution also has a Gaussian form:
\begin{equation}
\label{eq:maginal}
    \int {\text{d}{{\omega }_\text{i}} 
     {\left| F({{\omega }_\text{s}},{{\omega }_\text{i}}) \right|}^{2}} = {G(\omega_\text{s}|\omega_\text{p}/2, \Delta_+/2)},
\end{equation}
with the bandwidth twice narrower than $\Delta_+\equiv\sqrt{\Delta^2+\delta^2}\approx \Delta$.

%

\begin{subequations}\label{eq:Michelson1.PsiM}
The traditional MI-based autocorrelation QOCT scheme uses a detector that realizes a projection onto the two-photon state in mode $b$:
\begin{equation}\label{eq:Michelson1.PsiMbb}
    \left| {{\Psi }_{M}^{(bb)}}\left( {{\omega }_{1}},{{\omega }_{2}} \right) \right\rangle =\frac{1}{\sqrt{2}}\hat{b}_{\text{out}}^{\dagger }\left( {{\omega }_{1}} \right)\hat{b}_{\text{out}}^{\dagger }\left( {{\omega }_{2}} \right)\left| \text{vac} \right\rangle.
\end{equation}
As an alternative, we consider a cross-correlation scheme based on a detector that projects onto the state with two photons in different modes $a$ and $b$:
\begin{equation}\label{eq:Michelson1.PsiMab}
    \left| {{\Psi }_{M}^{(ab)}}\left( {{\omega }_{1}},{{\omega }_{2}} \right) \right\rangle =\hat{a}_{\text{out}}^{\dagger }\left( {{\omega }_{1}} \right)\hat{b}_{\text{out}}^{\dagger }\left( {{\omega }_{2}} \right)\left| \text{vac} \right\rangle.
\end{equation}
\end{subequations}
The probability to register one photon in the mode $m$ and another one in the mode $n$, 
as a function of the delay $\tau$ in the reference channel (the two-photon interferogram) is given by:
\begin{equation}
\label{eq:Michelson1.iterferogram1}
    M^{(mn)}\left( \tau  \right)=\iint{\text{d}{{\omega }_{1}}\text{d}{{\omega }_{2}}} {{\left| \left\langle  {{\Psi }_{M}^{(mn)}}\left( {{\omega }_{1}},{{\omega }_{2}} \right) | {{\Psi }_{\text{in}}} \right\rangle  \right|}^{2}}
    \end{equation}

Substituting Eqs.~(\ref{eq:Michelson1.matrices1}--\ref{eq:Michelson1.PsiM}) into Eq.~\eqref{eq:Michelson1.iterferogram1} we obtain the final interferograms for autocorrelation $(bb)$ and cross-correlation $(ab)$ measurement schemes:
\begin{subequations}\label{eq:Michelson1.FinalInterferogram}
\begin{align}
 \label{eq:Michelson1.FinalInterferogrambb}
   M_\text{MI}^\text{(auto)}&\equiv M^{(bb)}=\frac{M_c+M_0-M_1+M_2}{4 },\\
\label{eq:Michelson1.FinalInterferogramab}
   M_\text{MI}^\text{(cross)}&\equiv M^{(ab)}=\frac{M_c-M_0-M_2}{2}.
\end{align}
\end{subequations}
Here 
\begin{subequations}\label{eq:1M}
\begin{equation}
\label{eq:1.McM}{{M}_{c}}=\frac{1}{4}\left(1+R\right)^2    
\end{equation}
is a constant term. 
The term
\begin{equation}
   \label{eq:1.M0}{{M}_{0}}(\tau)=\frac{1}{2}R\exp \left[ -\frac{{{\left( T-\tau  \right)}^{2}}}{2{{\Delta }^{-2}}} \right]
\end{equation}
is related to HOM interference, presented in Fig.~\ref{fig:Schemes}b. 

The next term
\begin{equation}
   \label{eq:1.M1}{{M}_{1}}(\tau)=r\left( 1+R \right)\exp \left[ -\frac{{{\left( T-\tau  \right)}^{2}}}{8 \Delta _{+}^{-2}} \right]\cos \left[ \frac{{{\omega }_{\text{p}}}\left( T-\tau  \right)}{2} \right]
\end{equation}
is a term related to a single-photon Michelson interference. It fits up to a constant factor with the second term of the single-photon interferogram~\eqref{eq:classical} for the single-photon Gaussian spectral distribution~\eqref{eq:maginal}.
The last term
\begin{equation}
   \label{eq:1.M2}{{M}_{2}}(\tau)=\frac{1}{2}R\ \exp \left[ -\frac{{{\left( T-\tau  \right)}^{2}}}{2{{\delta }^{-2}}} \right]\cos \left[ {{\omega }_{\text{p}}}\left( T-\tau  \right) \right]
\end{equation}
\end{subequations}
is a similar term, related to Michelson interference of unseparated photon pairs, equivalent to the interference of narrowband pump photons.

The absolute values of Fourier-transforms of the time-dependent terms, presented in Fig.~\ref{fig:Schemes}c,~d are: 
\begin{subequations}
\begin{align}
    \abs{\tilde M_0(\omega)}&= \frac{\sqrt{\pi} R}{2} \ G(\omega|0,\Delta),\\
    \abs{\tilde M_1(\omega)}&= \frac{\sqrt{\pi} r (1+R)}{2}\ G\!\left(\omega\left|\frac{\omega_\text{p}}{2}\right.,\frac{\Delta_+}{2}\right),\\
    \abs{\tilde M_2(\omega)}&= \frac{\sqrt{\pi} R}{4}\ G(\omega, \omega_\text{p},\delta).  
\end{align}    
\end{subequations}

The target $M_0$ term can be easily extracted if it does not overlap with $M_1$ term in the Fourier domain, which requires $\Delta\ll\omega_\text{p}/3$. 
However, to achieve fine axial resolution ${\rm d}z\sim c/\Delta$ one needs to use a broadband SPDC source. 
For example, in Ref.~\cite{katamadze_broadband_2025} the autocorrelation MI-QOCT scheme with an SPDC source was used, covering the range 684--992~nm, which corresponds to $\Delta \approx 0.08\, \omega_\text{p}$. 
It resulted in non-negligible overlap between $\tilde M_0$ and $\tilde M_1$ peaks, which degraded the total QOCT resolution and SNR.

However, there is no $M_1$ term in the cross-correlation interferogram $M_\text{MI}^\text{(cross)}$. It has only one  
parasitic term $M_2$ with a spectrally separated peak and a narrow bandwidth $\delta$, which can be easily eliminated either by a Fourier-processing of the interferogram, or by averaging over the delay ${\rm d}\tau \sim 2\pi/\omega_\text{p}$ during a continuous delay scan. 

Additionally, as follows from Eq.~\eqref{eq:Michelson1.FinalInterferogram}, the target $M_0$ term in the interferogram $M_\text{MI}^\text{(cross)}$ has a twice the amplitude in comparison with $M_\text{MI}^\text{(auto)}$ and a negative sign, corresponding to the dip shape. 
The $M_1$ term in the interferogram $M_\text{MI}^\text{(auto)}$ has a positive sign, corresponding to a peak shape, since it is equivalent  to the probability to detect two photons in one output in the standard HOM-interference. 
Moreover, since usually two-photon probability is measured with a beam splitter and two single-photon detectors, as shown in Fig.~\ref{fig:Schemes}c, it leads to an additional 50\% loss in coincidence rate, thus the measured $M_1$ amplitude in the cross-correlation scheme is 4 times bigger than in the autocorrelation scheme.


\section{Experiment}\label{sec:Experiment}

To support our calculations, we conducted an experiment to benchmark the studied cross-correlation 
MI-QOCT against both the autocorrelation scheme and classical OCT, and to assess its potential applicability to real-world scenarios. The experimental setup is presented in Fig.~\ref{fig:setup}.
A degenerate SPDC source (see details in Appendix~\ref{appendix:source}) with a $50$~nm bandwidth and a central wavelength of $1313$~nm was coupled into a Michelson interferometer built with a fiberized beam-splitter (FBS). 
The reference and sample arms were taken out into free space. 
The mirror in the sample arm was translated with piezo or a stepper-motor stage to record interferograms. 

To record both cross-correlation $M_\text{MI}^\text{(cross)}(\tau)$ and autocorrelation $M_\text{MI}^\text{(auto)}(\tau)$ interferograms simultaneously, we inserted a fiber circulator at the interferometer input and added a second FBS at the output. 
Photon detection was performed with three near-infrared single‑photon detectors (ID Qube NIR), and coincidences were registered using a Time Tagger Ultra (Swabian Instruments).  
Autocorrelation interferograms were obtained from D2–D3 coincidences, while cross‑correlation interferograms were derived by summing D1–D2 and D1–D3 coincidences.
Additionally, single-photon interferograms, related to classical OCT, were obtained from the sum of D2 and D3 count rates.

\begin{figure}[!tbp]
    \centering
    \includegraphics[width = \columnwidth]{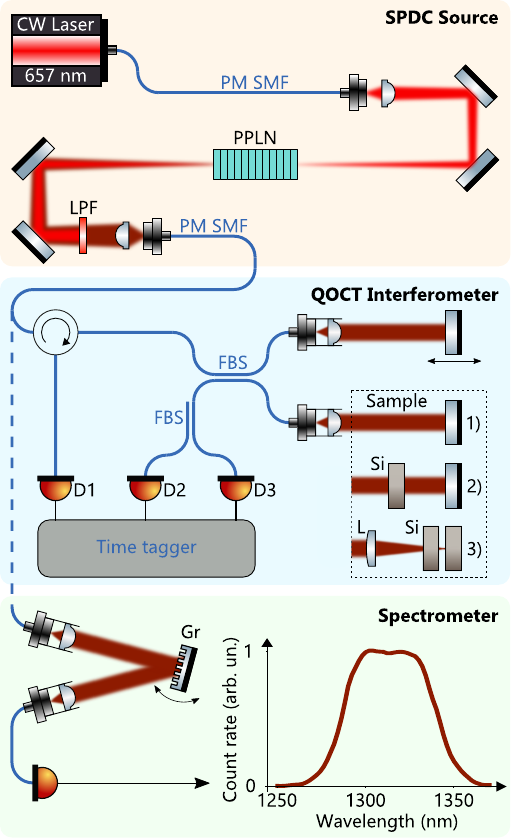}
    \caption{Experimental setup: A collinear degenerate SPDC Source (type-0 PPLN crystal pumped by a cw 657~nm laser) is coupled into a polarization‑maintaining single‑mode fiber and fed into a QOCT Interferometer built with fiber beam splitters (FBS) and two free‑space arms. A mirror in the reference arm is translated to record interferograms from the sample in the second arm. Samples include 1) a mirror,  2) a mirror with a silicon window, and 3) two silicon windows separated by an air gap. Three single‑photon detectors (D1, D2, D3), connected to a time tagger, measure both autocorrelation ($M_\mathrm{MI}^\mathrm{(auto)}$) and cross‑correlation ($M_\mathrm{MI}^\mathrm{(cross)}$) interferograms. The SPDC output is also sent to a home‑built spectrometer (two collimators, rotating grating Gr, and a single‑photon detector) to record the single‑count‑rate spectrum.}
     \label{fig:setup}
\end{figure}

In total, we conducted the following three experiments:
\begin{itemize}
  \item \textbf{Proof‑of‑concept:} A mirror sample was used to measure the PSF of the system to confirm the absence of the \(M_1\) term in cross‑correlation measurements.
  \item \textbf{Dispersion cancellation test:} A 5~mm‑thick silicon window was inserted before the mirror in the sample arm to demonstrate QOCT’s robustness against dispersion.
  \item \textbf{Defectoscopy application:} Two silicon windows served as a dispersive sample, and we resolved the air gap between them. To compensate for potential non‑parallelism of the window surfaces, the SPDC radiation was focused by lens~L. This also potentially allows transverse scanning of the sample, as in a 3D-imaging QOCT system.
\end{itemize}

\section{Experimental results and discussion}\label{sec:results}

\begin{figure*}[!p]
    \centering
    \includegraphics[width=\linewidth]{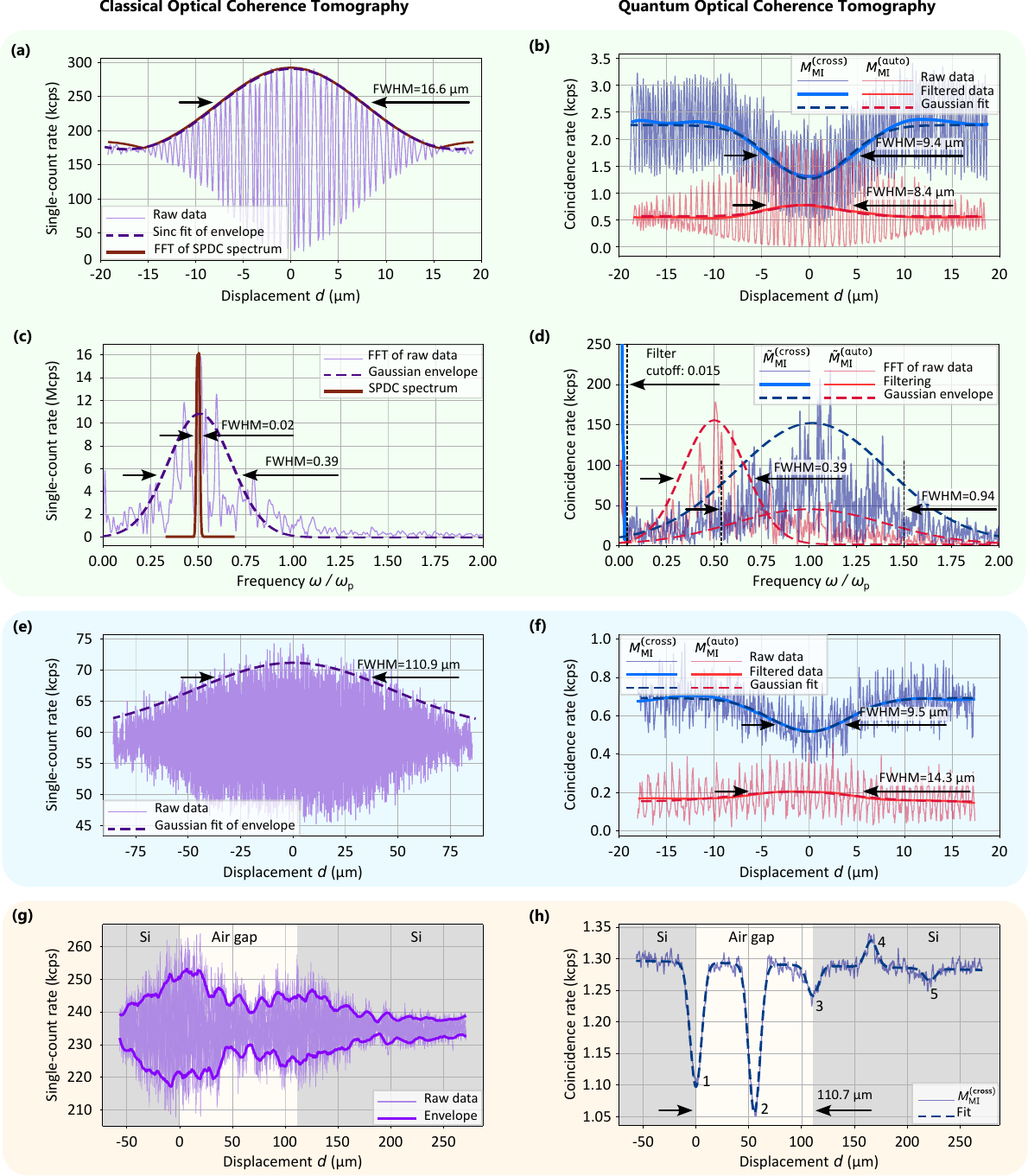}

\caption{Experimental results:  
(a) Single-count PSF interferogram for OCT: raw data (light-purple), envelope fit (thick dark-purple dashed), theoretical envelope (thick brown) obtained by inverse FFT of the SPDC source spectrum. 
(b) Coincidence-count PSF interferograms for QOCT: raw cross-correlation ($M_\mathrm{MI}^\mathrm{(cross)}$, thin light-blue) and autocorrelation ($M_\mathrm{MI}^\mathrm{(auto)}$, thin light-red), filtered HOM dips (thick red and blue), Gaussian fits (thick red and blue dashed). 
(c) Absolute FFT of the OCT interferogram in (a): raw (light-purple), envelope fit (thick dark-purple dashed), SPDC source spectrum (thick brown).
(d) FFTs of the QOCT data: solid blue and red curves, with darker segments indicating the filter region; dashed lines are Gaussian envelopes.
(e,f) Single-count OCT and coincidence-count QOCT PSF interferograms with dispersion introduced by a 5~mm silicon window. 
(g,h) Single-count OCT and coincidence-count QOCT interferograms for a sample of two silicon windows separated by a $\sim$110~\(\upmu\mathrm{m}\) air gap.}
    \label{fig:results}
\end{figure*}

\subsection{Proof-of-concept experiment}

The results of the proof-of-concept experiment are shown in the light-green block of Fig.~\ref{fig:results}.

The single-count interferogram, mimicking classical OCT, is presented in Fig.~\ref{fig:results}a as a thin light-purple solid line.  
To estimate the corresponding OCT resolution (PSF width), we fitted its envelope with a sinc function (thick dark-purple dashed curve) and obtained a value of $16.6 \pm 0.1\ \upmu$m.  
Here and below, uncertainties correspond to fitting errors.  

The FFT magnitude of this interferogram is shown in Fig.~\ref{fig:results}c.  
A Gaussian fit to its envelope (thick dark-purple dashed curve) yielded a spectral FWHM of $0.39 \pm 0.02\,\omega_\mathrm{p}$. 

For comparison, the directly measured SPDC spectrum (already shown in the bottom of Fig.~\ref{fig:setup}) is plotted as a brown line in Fig.~\ref{fig:results}c, and its inverse FFT is shown as the same line in Fig.~\ref{fig:results}a.  
We see that the interferogram envelope matches perfectly with the inverse FFT of the SPDC spectrum (as expected theoretically).  
However, the interferogram spectrum is nearly 20 times broader than the SPDC spectrum, indicating strong broadening due to phase noise in the fibers.  
Thus, although the SPDC spectrum fully determines the OCT resolution, the interferogram bandwidth is dominated by phase fluctuations.  

The raw QOCT coincidence interferograms $M_\mathrm{MI}^\mathrm{(cross)}(\tau)$ (light blue) and $M_\mathrm{MI}^\mathrm{(auto)}(\tau)$ (light red) are shown in Fig.~\ref{fig:results}b.  
These were recorded with a 70~nm step size and 100~ms exposure per point.  

Their FFT magnitudes are plotted in Fig.~\ref{fig:results}d with the same color coding.  
The envelopes were fitted with sums of Gaussian functions (dashed lines) centered at $\omega_\mathrm{p}/2$ and $\omega_\mathrm{p}$ (see details in Appendix~\ref{appendix:QOCTfit}).  
As expected, the autocorrelation spectrum $\tilde M_\mathrm{MI}^\mathrm{(auto)}$ exhibits both parasitic peaks $\tilde M_1$ and $\tilde M_2$.  
However, in contrast to the theory (Fig.~\ref{fig:Schemes}c), the $\tilde M_2$ peak is not narrow but instead twice as broad as $\tilde M_1$. This can be explained as follows: the same amplitude of path fluctuation produces different phase modulations at different wavelengths, resulting in a spectral broadening of $\tilde M_2$ that is roughly twice that of $\tilde M_1$.

These results, shown in Fig.~\ref{fig:results}d, demonstrate that even when the SPDC spectrum is relatively narrow and the parasitic peaks are not expected to overlap with the HOM term $\tilde M_0$, phase instability can still broaden them significantly, complicating digital spectral filtering.

By contrast, as expected, the $\tilde M_\mathrm{MI}^\mathrm{(cross)}(\tau)$ spectrum shows only the $\tilde M_2$ parasitic peak (the fitted amplitude of $\tilde M_1$ is negligible).

The $\tilde M_0$ terms are shown in Fig.~\ref{fig:results}d as thick blue and red curves.
To isolate them, we discarded FFT components above a cutoff frequency of $0.015\,\omega_\mathrm{p}$ and then applied an inverse FFT.
The resulting processed interferograms, plotted in Fig.\ref{fig:results}b with the same color coding, agree well with Gaussian fits (dashed curves).
The extracted FWHMs are $9.39 \pm 0.25\ \upmu$m and $8.42 \pm 0.35\ \upmu$m, respectively -- about half of the OCT PSF.
 
The small discrepancy likely arises because the single-count spectrum is broader than the coincidence-count spectrum due to coupling into the single-mode fiber~\cite{okanoDispersionCancellationHighresolution2013}.

Finally, the depth of the cross-correlation dip (995~kcps) exceeds the height of the autocorrelation peak (212~kcps) by more than a factor of four, demonstrating improved robustness against optical losses.

\subsection{Dispersion cancellation test}
The results of the dispersion cancellation experiment are shown in Fig.~\ref{fig:results}(e,f).  
To demonstrate the effect of the 5~mm silicon window on OCT, we first measured the single-count interferogram (OCT PSF) in Fig.~\ref{fig:results}e.  
Its envelope broadened by a factor of 6.68 to a FWHM of $110.9 \pm 6.8~\upmu$m. This value agrees with the calculated value of 6.64 in Eq.~\eqref{eq:Dispersion.broadening.OCT}, presented in Appendix~\ref{appendix:dispersion}.

The raw QOCT interferograms, recorded with a 70~nm step size and 200~ms exposure per point, are shown in Fig.~\ref{fig:results}f.  
We extracted the HOM terms as in the previous section.  
The FWHM of the processed QOCT PSFs remains nearly unchanged: $9.52 \pm 0.07~\upmu$m for cross‐correlation and $14.31 \pm 0.39~\upmu$m for autocorrelation.  
However, the silicon window introduces additional insertion losses, reducing the coincidence count rates to $0.175$~kcps (cross‐correlation) and $0.051$~kcps (autocorrelation), which made fitting the autocorrelation data challenging.  

These results confirm dispersion cancellation in both schemes; however, the cross‐correlation scheme is more robust to loss due to its higher SNR and simpler spectral filtering.  

To verify that the cross-correlation scheme’s SNR is suitable for real-world applications, we have conducted a demonstrative defectoscopy experiment, described below.

\subsection{Defectoscopy application}

In the previous experiments, we compared the two QOCT measurement schemes by performing sub‑wavelength, step‑scan measurements on a piezotranslator. This fine scanning allowed us to capture the full interferogram, including high‑frequency fringes, and then apply a digital low‑pass filter in post‑processing to isolate the low‑frequency HOM term. However, this method neither exploits the intrinsic phase‑noise insensitivity of HOM interference -- since it still requires sub‑wavelength path‑length control -- nor overcomes the piezo’s limited travel (100~µm).

For the defectoscopy application, we switched to hardware filtering. We drove a stepper‑motor translator continuously at 16.7~nm/s and binned photon counts into 300~nm intervals, which averages out the high‑frequency fringes in real time. Repeating each scan five times improved precision, and the stepper motor’s 25~mm range allowed us to interrogate the full thickness of the dispersive sample without sub‑wavelength positioning.

Figs~\ref{fig:results}(g,h) show the measured interferograms for an air gap between two silicon windows.  

Both OCT and QOCT interferograms were measured in parallel. To improve the SNR, the scan was repeated 8 times.

The single-photon OCT interferogram in Fig.~\ref{fig:results}g is severely blurred by dispersion, making layer separation impossible.  
In contrast, the QOCT coincidence interferogram in Fig.~\ref{fig:results}h exhibits distinct peaks and dips -- corresponding to reflections, echoes, and cross-interference terms~\cite{li-gomezQuantumEnhancedProbing2023} -- allowing us to resolve the window separation.  
Appendix~\ref{appendix:reconstruction} provides a detailed description of each enumerated term and the reconstruction algorithm.  
From this analysis, we extract a gap width of $110.67 \pm 0.11~\upmu\mathrm{m}$, in excellent agreement with the known spacer thickness.

\section{Conclusion}\label{sec:conclusion}

We have studied a cross‑correlation QOCT scheme based on a Michelson interferometer that overcomes the main limitations of previous implementations while preserving all of QOCT’s advantages. 
Compared with classical OCT, our QOCT scheme delivers a two-fold narrowing of the PSF and remains essentially immune to the sample dispersion that increases the classical OCT PSF by more than a factor of six.

Compared to the standard autocorrelation QOCT, the cross-correlation measurements achieve a fourfold improvement in SNR. 
Achieving high axial resolution requires a broadband light source. 
However, the autocorrelation scheme may not fully benefit from this, as it contains a parasitic peak at half the pump frequency, whose width increases with the source bandwidth. 
This peak overlaps with the HOM term, making it difficult to extract useful information.
A cross-correlation QOCT's interferogram spectrum contains only a single parasitic peak at the pump frequency -- well separated from the HOM peak -- which greatly simplifies extraction of the target HOM term by post-processing. This spectral separation reduces sensitivity to phase-instability-induced broadening and allows a compact, fiberized, vibration-insensitive implementation. Moreover, this spectral separation potentially enables combining our cross-correlation scheme with ultra-broadband SPDC sources to approach sub-wavelength axial resolution~\cite{katamadze_broadband_2025}. 

We have also shown that the HOM term can be directly recovered by averaging over the high-frequency oscillations during a continuous coarse mechanical scan, obviating the need for sub-wavelength path-length control and enabling long-range scanning.
Using this approach, we have shown that the SNR of our technique is high enough to localize a $110~\upmu\mathrm{m}$ air gap in a highly dispersive silicon medium.

Thus, the presented QOCT scheme enables the use of bright, broadband collinear type-0/type-I SPDC sources, is immune to optical dispersion and phase instability, and produces clear HOM dips with improved SNR -- paving the way to real-world QOCT applications.
\begin{acknowledgments}
Authors acknowledge useful discussions with Ren\'e Reimann, James Grieve, and  Rui Ming Chua.
\end{acknowledgments}

\appendix

\section{Single-photon and photon pair interference in a Michelson interferometer: general description}\label{appendix:MIgeneral}

Denote the sample response function as $H(\omega)$ and delay in the reference channel as $\tau$. The input and output modes of the interferometer are related according to Eqs.~(\ref{eq:Michelson1.matrices1}, \ref{eq:Michelson1.matrices2}).
Combining these equations, we obtain:
\begin{subequations}\label{eq:Michelson1.operators}
\begin{equation}
\hat{a}_{\text{in}}^{\dagger }(\omega)=\alpha(\omega)\hat{a}_{\text{out}}^{\dagger }(\omega)+\beta(\omega)\hat{b}_{\text{out}}^{\dagger}(\omega).
\end{equation}
were
\begin{equation}
\alpha(\omega)\equiv \frac{{{\operatorname{e}}^{\operatorname{i}\omega \tau }} + H\left( \omega  \right)}{2},\quad \beta(\omega)\equiv \ii \frac{{{\operatorname{e}}^{\operatorname{i}\omega \tau }} - H\left( \omega  \right)}{2}
.
\end{equation}
\end{subequations}

\subsection{Single-photon interference}
Consider a general case of a mixed single-photon state in the input mode $a$:
\begin{equation}\label{eq:SPin}
    \hat\rho_\text{in}=\iint{\text{d}\omega\text{d}\omega^\prime\ \rho(\omega, \omega^\prime)\ \hat a_\text{in}^\dagger(\omega)\ket{\text{vac}}\!\!\bra{\text{vac}}a_\text{in}(\omega^\prime),
    }
\end{equation}
were $\rho(\omega, \omega^\prime)$ is a cross-spectral density function.

Single-photon measurement in the mode $b$ corresponds to a projection to a state:
\begin{equation}\label{eq:psim}
    \ket{\psi_M^{(b)}(\omega)}=\hat b_\text{out}(\omega)\ket{vac}.
\end{equation}

Then the total probability to detect photon in the output mode $b$ depending on the delay $\tau$ gives the interferogram:
\begin{equation}\label{eq:MOCT1}
    M^{(b)}(\tau) = \int{\text{d}\omega \bra{\psi_M^{(b)}(\omega)}\hat\rho_\text{in}\ket{\psi_M^{(b)}(\omega)}}.
\end{equation}

Substituting Eqs.~(\ref{eq:Michelson1.operators}--\ref{eq:psim}) into Eq.~\eqref{eq:MOCT1} we obtain:

\begin{equation}\label{eq:MOCT2}
    M^{(b)}(\tau) = \int{\text{d}\omega\  \rho(\omega,\omega)\abs{\beta(\omega)}^2
    }.
\end{equation}
Note that Eq.~\eqref{eq:MOCT2} depends on the diagonal part of the crossed-spectral density function only, which describes the spectral distribution of the state and does not depend on its coherence and purity. So, the same interferogram could be obtained with coherent or thermal light used in classical OCT setups. 

For the Gaussian spectral distribution \eqref{eq:gaussian} and the single-layer response function \eqref{eq:single} the Eq.~\eqref{eq:MOCT2} transforms to Eq.~\eqref{eq:classical}.

\subsection{Photon-pair interference}
Consider a photon-pair input state \eqref{eq:Michelson1.Psiin} in the input mode $a$. 
Assume the spectral amplitude $F\left( {{\omega }_\text{s}},{{\omega }_\text{i}} \right)$ is symmetrical, so that $F\left( {{\omega }_\text{s}},{{\omega }_\text{i}} \right)=F\left( {{\omega }_\text{i}},{{\omega }_\text{s}} \right)$, which is true for type-I or type-0 SPDC.

Substituting \eqref{eq:Michelson1.operators} into \eqref{eq:Michelson1.Psiin} we obtain
\begin{equation}
\begin{aligned}\label{eq:Michelson1.Psiin1}
&\left| {{\Psi }_{\text{in}}} \right\rangle=\frac{1}{{\sqrt{2}}}\times\\
\Bigg[&
\iint{\text{d}{{\omega }_\text{s}}\text{d}{{\omega }_\text{i}}}
F_{\text{out}}^{(aa)}\left( {{\omega }_\text{s}},{{\omega }_\text{i}} \right)
    \hat a^\dagger_{\text{out}}(\omega_{\text{s}})
    \hat a^\dagger_{\text{out}}(\omega_{\text{i}})
    \left| \text{vac} \right\rangle +\\
&
\iint{\text{d}{{\omega }_\text{s}}\text{d}{{\omega }_\text{i}}}
F_{\text{out}}^{(ab)}\left( {{\omega }_\text{s}},{{\omega }_\text{i}} \right)
    \hat a^\dagger_{\text{out}}(\omega_{\text{s}})
    \hat b^\dagger_{\text{out}}(\omega_{\text{i}})
    \left| \text{vac} \right\rangle +\\
&
\iint{\text{d}{{\omega }_\text{s}}\text{d}{{\omega }_\text{i}}}
F_{\text{out}}^{(bb)}\left( {{\omega }_\text{s}},{{\omega }_\text{i}} \right)
    \hat b^\dagger_{\text{out}}(\omega_{\text{s}})
    \hat b^\dagger_{\text{out}}(\omega_{\text{i}})
    \left| \text{vac} \right\rangle\Bigg]
\end{aligned}
\end{equation}
were
\begin{subequations}\begin{align}\label{eq:Michelson1.Psiin2}
F_{\text{out}}^{(aa)}\left( {{\omega }_\text{s}},{{\omega }_\text{i}} \right)\equiv&
\alpha(\omega_\text{s})\alpha(\omega_\text{i}) F\left( {{\omega }_\text{s}},{{\omega }_\text{i}} \right) \\
F_{\text{out}}^{(ab)}\left( {{\omega }_\text{s}},{{\omega }_\text{i}} \right)\equiv&2
\alpha(\omega_\text{s})\beta(\omega_\text{i})F\left( {{\omega }_\text{s}},{{\omega }_\text{i}} \right)\\
\nonumber
F_{\text{out}}^{(bb)}\left( {{\omega }_\text{s}},{{\omega }_\text{i}} \right)\equiv&
\beta(\omega_\text{s})\beta(\omega_\text{i})F\left( {{\omega }_\text{s}},{{\omega }_\text{i}} \right)
\end{align}
\end{subequations}

Here the upper index $(mn)$ is related to the state with one photon in $m$ mode and another one in $n$ mode.

The projections for autocorrelation and cross-correlation measurements are given by 
\begin{subequations}
\begin{equation}
\begin{aligned}
  &\left\langle  {{\Psi }^{(bb)}_{M}}\left( {{\omega }_{1}},{{\omega }_{2}} \right) | {{\Psi }_{\text{in}}} \right\rangle 
=\\
 & \quad \frac{1}{2}\left[ {F^{(bb)}_{\text{out}}}\left( {{\omega }_{1}},{{\omega }_{2}} \right)+{F^{(bb)}_{\text{out}}}\left( {{\omega }_{2}},{{\omega }_{1}} \right) \right]=\\
 &\quad\beta(\omega_1)\beta(\omega_2)F(\omega_1,\omega_2),
\end{aligned}
\end{equation}
\begin{equation}
\begin{aligned}
  &\left\langle  {{\Psi }^{(ab)}_{M}}\left( {{\omega }_{1}},{{\omega }_{2}} \right) | {{\Psi }_{\text{out}}} \right\rangle = 
\frac{1}{\sqrt{2}} {F}^{(ab)}_{\text{out}}\left( {{\omega }_{1}},{{\omega }_{2}} \right) =\\
& \quad
\frac{1}{\sqrt{2}}\alpha(\omega_1)\beta(\omega_2)F(\omega_1,\omega_2)
\end{aligned}
\end{equation}
\end{subequations}
respectively.

Let us make a change of variables: $\omega_1=\omega_0+\nu_1$, $\omega_2=\omega_0+\nu_2$, where $\omega_0\equiv\omega_{\text{p}}/2$ -- half of the pump frequency.
In this case, we have
\begin{subequations}
\begin{equation}
\begin{aligned}
&\abs{\beta(\omega_0+\nu_1)\beta(\omega_0-\nu_2)}^2 =\\
&\quad\quad K_c+K_0-(K_1+K_{1^\prime})/2+K_2,
\end{aligned}
\end{equation}
\begin{equation}
\begin{aligned}
&\abs{\alpha(\omega_0+\nu_1)\beta(\omega_0-\nu_2)}^2 =\\
 &\quad\quad K_c-K_0+(K_1 - K_{1^\prime})/2 - K_2,
\end{aligned}
\end{equation}
\end{subequations}
where
\begin{subequations}\label{eq:Michelson1.K}
\begin{align}
K_c\equiv &\frac{1}{4}\left(\abs{H(\omega_0+\nu_1)}^2+1\right)\left(\abs{H(\omega_0+\nu_2)}^2+1\right)\label{eq:Michelson1.Kc} ,\\
K_0\equiv & \frac{1}{2}\Re\left[ \ee^{-\ii(\nu_1-\nu_2)\tau} H(\omega_0+\nu_1)H^\ast(\omega_0+\nu_2) \right]\label{eq:Michelson1.K0},\\ \nonumber
K_1\equiv &\nonumber \\\Re&\left[\ee^{-\ii\omega_0\tau} \ee^{-\ii\nu_1\tau}H(\omega_0+\nu_1)\left(\abs{H(\omega_0+\nu_2)}^2+1\right)\right]
\label{eq:Michelson1.K1},\\
K_{1^\prime}\equiv &\nonumber \\
\Re&\left[\ee^{-\ii\omega_0\tau} \ee^{-\ii\nu_2\tau}H(\omega_0+\nu_2)\left(\abs{H(\omega_0+\nu_1)}^2+1\right)\right]\label{eq:Michelson1.K1prime},\\
K_2\equiv& \frac{1}{2}\Re\left[\ee^{-2\ii\omega_0\tau}\ee^{-\ii(\nu_1+\nu_2)\tau}H(\omega_0+\nu_1)H(\omega_0+\nu_2)
\right].\label{eq:Michelson1.K2}
\end{align}
\end{subequations}
Therefore the final interferograms consist of five terms:
\begin{subequations}\label{eq:Michelson1.FinalInterferogramApp}
\begin{equation}\label{eq:Michelson1.FinalInterferogram1bbApp}
   4 M^{(bb)}(\tau)=M_c^{(M)}+M_0(\tau)-\frac{M_1(\tau)+M_{1^\prime}(\tau)}{2}+M_2(\tau),
\end{equation}
\begin{equation}\label{eq:Michelson1.FinalInterferogram1abApp}
   2 M^{(ab)}(\tau)=M_c^{(M)}-M_0(\tau)+\frac{M_1(\tau)-M_{1^\prime}(\tau)}{2}-M_2(\tau),
\end{equation}
\end{subequations}
where
\begin{equation}\label{eq:Michelson1.Mj}
    {{M}_{j}}=\iint{\text{d}{{\nu }_{1}}\text{d}{{\nu }_{2}}}{{\left| F\left( {{\omega }_{0}}+{{\nu }_{1}},{{\omega }_{0}}+{{\nu }_{2}} \right) \right|}^{2}}{K}_{j}
\end{equation}
for $j=c,0,1,1^\prime,2$. It is easy to see that $M_1(\tau)=M_{1^\prime}(\tau)$, therefore we obtainEqs.~\eqref{eq:Michelson1.FinalInterferogram} for the interferograms for both measurement schemes. Substituting the Gaussian photon pair spectral distribution \eqref{eq:degenerate} into Eq.~(\ref{eq:Michelson1.Mj}) we get the final form of the interferogram terms~\eqref{eq:1M}.


\section{SPDC source}\label{appendix:source}

The SPDC part of the experiment employed a 656.5~nm CW pump laser (linewidth $\sim 0.01$~nm) to generate photon pairs via type-0 SPDC in a 25 mm periodically-poled lithium niobate (PPLN) crystal with a 12.83 $\upmu$m poling period (HC Photonics) at a stabilized temperature of 50$^\circ\text{C}$. The pump light was PM fiber-coupled and focused into the crystal using an aspheric lens, while the SPDC light was collected into telecom fiber using another aspheric lens. The pump and collection focal parameters \cite{dixon_heralding_2014} $\xi_\text{p} = 0.35$ and  $\xi_\text{c} = 1$ were chosen as a trade-off between heralding efficiency and source brightness \cite{beckert_space-suitable_2019}, 
and their values were verified using CCD cameras to measure the beam waists at the PPLN location.

\section{Dispersion Broadening}\label{appendix:dispersion}

The effect of dispersion on OCT and QOCT resolution was analyzed in Ref.~\cite{katamadze_broadband_2025}.  From Eqs.~(16b,c) and (21b,c) of its Supplemental Material, the FWHM of the HOM term \(M_0(\tau)\), which determines the QOCT axial resolution, increases by a factor
\[
\alpha_0 \;=\;\sqrt{1 + \delta^2 \,\Delta^2 \,\kappa^2}\,,
\]
while the FWHM of the single‐photon Michelson term \(M_1(\tau)\), which corresponds to the classical OCT resolution, increases by
\[
\alpha_1 \;=\;\sqrt{1 + \frac{\Delta_+^4\,\kappa^2}{4}}\,.
\]
Here
\[
\kappa \;=\; \frac{l}{c}\,\left.\frac{\partial n}{\partial \omega}\right|_{\omega_p/2},
\]
with \(n(\omega)\) the refractive index of the dispersive medium and \(l\) the total round‐trip path length through that medium.

Using the Sellmeier equation for silicon from Ref.~\cite{tatian_fitting_1984}, one finds
\[
\left.\frac{\partial n}{\partial \omega}\right|_{\omega_p/2}
\approx -1.3 \times 10^{-16}\,\mathrm{s}.
\]
For our experimental parameters -- a pump linewidth of 0.01~nm (\(\delta = 2\pi\times6.9\)~GHz), an SPDC bandwidth of 50~nm (\(\Delta = 2\pi\times8.7\)~THz), and a silicon plate thickness \(l/2 = 5\)~mm -- this yields

\begin{subequations}\label{eq:Dispersion.broadening}
\begin{equation}\label{eq:Dispersion.broadening.QOCT}
   \alpha_0 \approx 1 + 5.5\times10^{-5} \approx 1,
\end{equation}
\begin{equation}\label{eq:Dispersion.broadening.OCT}
   \alpha_1 \approx \sqrt{1 + \frac{\bigl( \delta^2 + \Delta^2 \bigr)^2 \kappa^2}{4}} \approx 6.64.
\end{equation}
\end{subequations}

Thus, a 5~mm silicon window negligibly affects the QOCT resolution but broadens the classical OCT point‐spread function by more than a factor of six.

\section{Data processing details}

\subsection{QOCT spectra fit}\label{appendix:QOCTfit} 
Since in general the QOCT interferogram spectra contain two parasitic terms, $\tilde M_1$ and $\tilde M_2$, centered at $\omega_\text{p}/2$ and $\omega_\text{p}$ respectively, we approximate  
$\tilde M_\mathrm{MI}^\mathrm{(cross)}$ and $\tilde M_\mathrm{MI}^\mathrm{(auto)}$ (Fig.~\ref{fig:results}d) by the sum of Gaussian functions:  
\begin{equation}
     f_{\text{fit}}(\omega) = c + a_1 \exp\left[-\frac{(\omega - \omega_\text{p}/2)^2}{2\sigma_1^2}\right] + a_2 \exp\left[-\frac{(\omega - \omega_\text{p})^2}{2\sigma_2^2}\right].
\end{equation}

\begin{figure}[!tbp]
    \centering
    \includegraphics[width = \columnwidth]{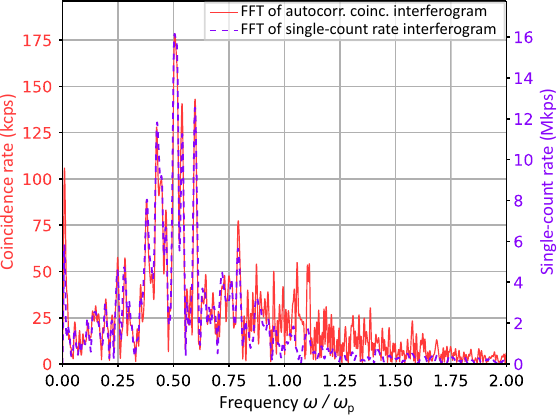}
    \caption{
     Overlap of interferogram spectra from Fig.~\ref{fig:results}(c, d): FFT magnitude of the QOCT coincidence-rate autocorrelation interferogram $\tilde M_\mathrm{MI}^\mathrm{(auto)}$ (solid red) and FFT magnitude of the OCT single-count interferogram $\tilde M_\mathrm{OCT}$ (dashed purple).
    }
     \label{fig:single_coinc}
\end{figure}

Because the single-count spectrum and the part of $\tilde M_\mathrm{MI}^\mathrm{(auto)}$ around $\omega_\text{p}/2$ are both affected by the same phase fluctuations and therefore coincide (Fig.~\ref{fig:single_coinc}), we fixed the standard deviation of the first Gaussian to $\sigma_1 = 0.16 \,\omega_\text{p}$, as extracted from the Gaussian fit of the single-count spectrum (Fig.~\ref{fig:results}c, dark-purple dashed curve).  
Table~\ref{tab:gaussian_fit} summarizes the extracted values of all other parameters with fitting errors.

\begin{table}[h]
  \centering
  \caption{Parameters of Gaussian fit for $\tilde M_\mathrm{MI}^\mathrm{(cross)}$ and $\tilde M_\mathrm{MI}^\mathrm{(auto)}$}
  \begin{tabular}{c|l|c|r|r}
    Parameter & 
       \multicolumn{1}{c|}{$\tilde M_\mathrm{MI}^\mathrm{(cross)}$}
      & $\tilde M_\mathrm{MI}^\mathrm{(auto)}$ \\ \hline
    $c$ (kcps) &  4.6 $\pm$ 1.8 &  4.6 $\pm$1.3 \\
    $a_1$ (kcps) & $1.2\times10^{-13} \pm 10$&  112$\pm$ 7 \\
    $a_2$ (kcps)& 148$\pm$ 6 & 39  $\pm$4  \\
    $\sigma_2$ ($\omega_\text{p}$) & 0.40 $\pm$0.02 & 0.40$\pm$ 0.05\\
  \end{tabular}
  \label{tab:gaussian_fit}
\end{table}

The amplitude $a_1$ extracted for $\tilde M_\mathrm{MI}^\mathrm{(cross)}$ is negligibly small ($10^{-13} \pm 10$ kcps), confirming the absence of the $\tilde M_1$ peak.  
The fitted values of $\sigma_2$ for the $\tilde M_2$ peaks in both $\tilde M_\mathrm{MI}^\mathrm{(auto)}$ and $\tilde M_\mathrm{MI}^\mathrm{(cross)}$ are $0.40\,\omega_\text{p}$, nearly twice as large as $\sigma_1$.  
This agrees with the expectation that the same optical path fluctuation amplitude induces twice the phase modulation depth at twice the frequency, leading to proportionally larger spectral broadening.

\subsection{OCT and QOCT data processing for defectoscopy experiment}

During the defectoscopy experiment, we repeated the scan eight times.  
Because the source intensity was not perfectly stable within and between runs, we introduced a normalization procedure
 to eliminate slow drifts in brightness and make the different traces comparable.  
Each single-count and coincidence trace was normalized by  
\begin{equation}
    R_\text{norm} = \frac{R_1 + \kappa_2 R_2 + \kappa_3 R_3}{\left< R_1 + \kappa_2 R_2 + \kappa_3 R_3 \right>}, 
\end{equation}
where $R_1$, $R_2$, and $R_3$ are the single-count rates from detectors D1, D2, and D3, respectively; $\langle \dots \rangle$ denotes averaging over the entire trace; and the coefficients $\kappa_2$, $\kappa_3$ were chosen such that $\langle R_1 \rangle = 2\kappa_2 \langle R_2 \rangle = 2\kappa_3 \langle R_3 \rangle$.

All single-count interferograms are shown in Fig.~\ref{fig:results}g.  
To align the single-count traces from different runs, we subtracted the mean and divided by the standard deviation for each trace, then rescaled by adding back the global average mean and multiplying by the global average standard deviation.  
To obtain an approximate envelope, we determined at each point the maximum and minimum values across all measurements (upper and lower envelopes) and then smoothed both curves with a Savitzky–Golay filter (window $\sim$17~$\upmu$m, third-order polynomial).  

The cross-correlation interferogram $\tilde M_\mathrm{MI}^\mathrm{(cross)}$ (light-blue curve in Fig.~\ref{fig:results}h) was obtained by averaging the coincidence data across all eight measurements.

\subsection{Reconstruction of the sample structure}\label{appendix:reconstruction}
 
Following Ref.~\cite{li-gomezQuantumEnhancedProbing2023}, a QOCT interferogram comprises three types of terms (dips or peaks) with amplitudes \(V\) (we define \(V>0\) for dips and \(V<0\) for peaks):

\begin{itemize}
  \item \textbf{Direct reflections.}  
    Reflections from each interface appear at 
    \[
      x_1 = 0~\upmu\mathrm{m}, 
      \quad
      x_2 = x_1 + d,
    \]
    with amplitudes
    \begin{align}
      V_1 &= \frac{A}{2}\bigl|\tilde r_1\bigr|^2, 
      &\tilde r_1 &= r_1,\\
      V_2 &= \frac{A}{2}\bigl|\tilde r_2\bigr|^2, 
      &\tilde r_2 &= r_1\,r_2\,t_1^2,
    \end{align}
    where \(A\) is a common amplitude factor, \(r_1\) and \(r_2\) are the complex reflection coefficients of the first and second interfaces, \(t_1\) is the complex transmission coefficient of the first interface, and \(d\) is the interface separation.

  \item \textbf{Echoes.}  
    Multiple internal reflections produce echoes at
    \[
      x_k = x_1 + k\,d,\quad k=2,3,4,\dots
    \]
    with visibility
    \begin{equation}
      V_k = \frac{A}{2}\bigl|\tilde r_k\bigr|^2,\quad
      \tilde r_k = r_1\,r_2\,r_{k-1}.
    \end{equation}

  \item \textbf{Cross terms.}  
    Every pair of dips at \(x_k\) and \(x_l\) generates a cross term at
    \(\displaystyle x_{kl} = \frac{x_k + x_l}{2}\)
    with amplitude
    \begin{equation}
      V_{kl} = A\,\mathrm{Re}\bigl[\tilde r_k\,\ee^{\ii\omega_0 k d/c}\,\tilde r_l^*\,\ee^{-\ii\omega_0 l d/c}\bigr].
    \end{equation}
    Unlike the previous terms, \(V_{kl}\) can be negative (peaks).  
    We omit the pump coherence–length factor since it significantly exceeds the interferogram length.
\end{itemize}

The interferogram of Fig.~\ref{fig:results}h was first fitted with the following function:
\begin{equation}\label{eq:fitfunc}
    f_{\text{fit}}(x) = C - \sum\limits_{j=1}^5 A_j \exp\left[- \frac{(x - x_1 - jd/2)^2}{2\sigma^2}\right] + kx,
\end{equation}
with the fitting parameters: $A_1,\dots,A_5$ (enumeration corresponds to the term numbers in Fig.~\ref{fig:results}h), $\sigma$, $x_1$, $k$ and the desired separation $d$.

\begin{table}[h]
  \centering
  \caption{Experimental amplitudes of the terms in Fig.~\ref{fig:results}h, and their theoretical approximation.}
  \label{tab:fit}
  \begin{tabular}{c|l|c|r|r}
    & 
      & \multicolumn{1}{c|}{Theoretical} & \multicolumn{1}{c|}{Exper.} & \multicolumn{1}{c}{Approx.} \\
   
     Amplitude& \multicolumn{1}{c|}{Position} & \multicolumn{1}{c|}{approximation} & \multicolumn{1}{c|}{value} & \multicolumn{1}{c}{value} \\
      &
      & &\multicolumn{1}{c|}{(cps)} & \multicolumn{1}{c}{(cps)} \\ \hline
    $A_1$ & \(x_1\) & \(V_1\)                & $196\pm 2$ & 196 \\
    $A_2$ & \(x_1 + {d}/{2}\) & \(V_{12}\)           & $237\pm 2$ & 237 \\
    $A_3$ & \(x_1+d\) & \(V_2 + V_{13}\)         & $50\pm 2$ &  50 \\
    $A_4$ & \(x_1+ {3d}/{2}\) & \(V_{23} + V_{14}\)   &$-38\pm2$ & $-40$ \\
    $A_5$ & \(x_1 + 2 d\) & \(V_3 + V_{24} + V_{15}\)& $20\pm 2$ &  17 \\
  \end{tabular}
\end{table}

The amplitudes $A_1,\dots,A_5$ were then approximated by the theoretical expressions in Table~\ref{tab:fit}, where the free parameters \(A\), \(r_1\), \(r_2\), \(t_1\), and \(\phi_0 \equiv \omega_0 d/c\) were adjusted so as to minimize the discrepancy between experiment and theory.

The approximation shows a good agreement with the experimental data. 
Finally, fixing the amplitudes to their approximated values, we refitted the interferogram terms to extract the interface separation,
  $d = 110.67 \pm 0.11~\upmu\mathrm{m}$.

\bibliography{apssamp}

\end{document}